\documentclass[12pt,a4paper,dvips]{article}

\usepackage{graphicx}
\usepackage{amsbsy}
\usepackage{amssymb}

\begin{document}

\newcommand{\be}{\begin{equation}}
\newcommand{\ee}{\end{equation}}
\newcommand{\bea}{\begin{eqnarray}}
\newcommand{\eea}{\end{eqnarray}}
\newcommand{\tr}{\,\hbox{tr }}
\newcommand{\Tr}{\,\hbox{Tr }}
\newcommand{\Det}{\,\hbox{Det }}
\newcommand{\fslash}{\hskip-.2cm /}
\title{A Non-Singular Black Hole}

\author{
Aleksandar Bogojevi\'c
\thanks{http://www.phy.bg.ac.yu/ctp/pft/a\_bogojevic/}\\
\textsl{Institute of Physics}\\
\textsl{P.O.B. 57, Belgrade 11001}\\
\textsl{Yugoslavia}\\
\\
and\\
\\
Dejan Stojkovi\'c\\
\textsl{Department of Physics}\\
\textsl{Case Western Reserve University}\\
\textsl{Cleveland, OH 44106}
}

\date{Preprint IP-HET-98/10\\
April 1998}
\maketitle

\begin{abstract}
We present a completely integrable deformation of the CGHS 
dilaton gravity model in two dimensions. The solution is a singularity
free black hole that at large distances asymptoticaly joins to the
CGHS solution.
\end{abstract}
\newpage

\section{Introduction}

One of the fundamental unsolved problems in theoretical physics is the
unification of quantum theory and gravity. Many reason why this has
proved so difficult stem from the complicated nonlinear structure of the
equations of general relativity. Gravitational equations are much simpler
in lower dimensions. This is the reason why there has been so much activity
related to the quantization of gravity in two and three dimensions.
\cite{teitelboim}-\cite{fis}. One of the most important results in 2d was
the exactly solvable dilaton gravity model constructed by Callan, Giddings,
Harvey and Strominger. The CGHS model has 2d black hole solutions that are
remarkably similar to the Schwarzschild solution of general relativity.

Of the four fundamental interactions in nature, gravity is by far the
weakest \cite{hs}-\cite{mann}. For this reason, we can hope to see quantum
effects only in the vicinity of classical singularities. Penrose and Hawking
\cite{penrose}-\cite{hp} have shown that these singularities are endemic in
general relativity. The general belief is that quantization will rid
gravitation of singularities, just as in atomic physics it got rid of the
singularity of the Coulomb potential. If this is indeed the case, then there
must exist a non-singular gravitational effective action whose classical
equations encode the full quantum theory. This effective action must have
the Planck length $L_\mathrm{Planck}$ in it as an input parameter. For
$L\gg L_\mathrm{Planck}$ the effective model must be indistinguishable from
the classical gravity action. In an interesting series of papers
\cite{mb}-\cite{tmb} Brandenberger, Mukhanov and their collaborators have
initiated a search for such effective models of 2d dilaton gravity. They
investigated a proceedure by which one could make models free of
singularities. This parallels Landau's treatment of phase transitions in
ferromagnets. Landau chose (the simplest) effective action (Gibbs potential
in statistical mechanics parlance) that led to a qualitatively correct
discription of phase transitions.

A recent success in the field of 2d dilaton gravity has been the work of
Louis-Martinez and Kunstatter \cite{lmk}, who reduced the solution of the
general dilaton gravity model to the solution of two ordinary integrals,
i.e. to two quadratures. In this paper we will use their result to
construct an exactly solvable class of models --- deformed CGHS models. For
$L\gg L_\mathrm{Planck}$ these models go over into the CGHS model. Like
CGHS, the deformed models are exactly solvable (the two quadrature integrals
can be calculated in closed form) and have black hole solutions (solutions
with event horizons). Unlike CGHS, the deformed models are non-singular.
As we shall see, the maximal curvature is proportional to
$\frac{1}{L_\mathrm{Planck}}$.

\section{CGHS Model}

The action of all dilaton gravity models can be put into the general form
\be
\label{general action}
S=\int d^2x \sqrt{-g}\left[\,{1\over 2}\,g^{\alpha\beta} 
\partial_\alpha \phi 
\partial_\beta \phi - V(\phi )+D(\phi)R\right]\ .
\ee
The potentials $V(\phi)$ and $D(\phi)$ classify all the possible models. 
Let us perform a conformal scaling of the metric
\be 
\tilde g_{\alpha\beta}=e^{-2F(\phi)}g_{\alpha\beta}\ ,
\ee
where the scaling factor $F(\phi)$ satisfies
\be 
\label{first quadrature}
{{dF}\over {d\phi}}=-{1\over 4} \; 
\left( {{dD}\over {d\phi}}\right) ^{-1}\ .
\ee
This puts the action into the simplified form
\be 
\label{simplified action}
S=\int d^2 x\sqrt {-\tilde g}
\left[ \tilde\phi \tilde R-\tilde V (\tilde \phi)\right]\ ,
\ee
where $\tilde R$ is the scalar curvature corresponding to 
$\tilde g_{\alpha\beta}$, and we have introduced the new dilaton field and
potential according to
\bea
\label{phi tilde}
\tilde \phi&=&D(\phi)\\
\tilde V(\tilde\phi)&=&e^{2F(\phi)}V(\phi)\ .
\eea
This form of the dilaton gravity action is obviously much easier to work
with since we have lost the kinetic term for the dilaton field.

A well known property of two dimensional manifolds allows us to localy,
i.e. patch by patch, choose conformally flat coordinates for which
\be
\tilde g_{\alpha\beta}=e^{2\rho}\,\eta _{\alpha \beta}\ .
\ee
Louis-Martinez and Kunstatter \cite{lmk} have shown that we can choose a
coordinate system in which the solution of the general dilaton model
is static and given by
\bea
\label{second quadrature} 
x&=&-2\int {d\tilde\phi\over W(\tilde\phi)+C}\\
e^{2\rho}&=&-\,{C+W(\tilde\phi)\over 4}\ ,
\eea
where the pre-potential $W(\tilde\phi)$ is given by 
${dW\over d\tilde\phi}=\tilde V(\tilde\phi)$, and $C$ is an invariant.
As we can see, the solution is given in terms of two quadratures: equations
(\ref{first quadrature}) and (\ref{second quadrature}), determining
$F(\phi)$ and $\tilde\phi(x)$ respectively. A given model is completely
integrable only if we can calculate both quadratures in closed form.

The CGHS model is an example of a completely integrable dilaton gravity
model. The standard form of the CGHS action is
\be
S=\int d^2x\sqrt{-g}\,e^{-2\varphi}\,(R+4g^{\alpha\beta}
\partial_\alpha\varphi\partial_\beta\varphi +4\lambda^2)\ .
\ee
The simple field redefinition $\phi=\sqrt{8}\, e^{-\varphi}$ puts this into
the general form for dilaton gravity actions given in (\ref{general action}).
We find
\be 
\label{CGHS}
S=\int d^2x\sqrt {-g}
\left(\,{1\over 2}\,g^{\alpha\beta}
\partial_\alpha\phi\partial_\beta\phi+
{1\over 2}\,\lambda^2\phi^2+{1\over 8}\,R\phi^2\right)\ ,
\ee
hence
\bea
V(\phi)&=&-{1\over 2}\,\lambda^2\phi^2\\
D(\phi)&=&{1\over 8}\,\phi^2\ .
\eea
The first quadrature is easily integrated and we get
\be
F(\phi)=-\ln\phi\ .
\ee
Using this, $\tilde\phi=D(\phi)$, as well as the definition of 
$\tilde V(\tilde\phi)$ we immediately find that 
$\tilde V(\tilde\phi)=-{1\over 2}\lambda^2$, and thus the pre-potential is
$\tilde W(\tilde\phi)=-{1\over 2}\lambda^2\tilde\phi$. 
The simplified form of the CGHS action is therefore
\be
S=\int d^2x\sqrt{-\tilde g}
\left(\tilde\phi\tilde R+{1\over 2}\,\lambda^2\right)\ .
\ee

The CGHS model is completely integrable. In our notation this means that the
second quadrature (\ref{second quadrature}) can also be solved in closed
form. A trivial integration gives
\be
x={4\over\lambda^2}\ln
\left(\,{1\over 2}\,\lambda^2\tilde\phi-C\right)\ .
\ee
Inverting this we find
\be
\label{phi tilde x}
\tilde\phi(x)={2\over\lambda^2}
\left(e^{{\lambda^2\over 4}x}+C\right)\ .
\ee
According to the general prescription this gives
\bea 
\label{rho}
\rho(x)&=&-\ln 2+{\lambda^2\over 8}x\\
\phi(x)&=&{4\over\lambda}
\left(e^{{\lambda^2\over 4}x}+C\right)^{1\over 2}\ .
\eea
This, along with the expression for $F(\phi)$, gives us
\be
F(x)=-\ln{4\over\lambda}
-{1\over 2}\ln\left(e^{{\lambda^2\over 4}x}+C\right)\ .
\ee
The scalar curvature of the general model can be given in terms of
$\rho(x)$ and $F(x)$. We find
\be
\label{curvature}
R=-2\,e^{-2(F+\rho)}{d^2\over dx^2}(F+\rho)\ .
\ee
For the CGHS model this gives
\be
R={4\lambda^2C\over e^{{\lambda^2\over 4}x}+C}\ .
\ee
Obviously $R$ has a singularity for $C<0$. This is the CGHS black hole
solution. In fact, it can be shown that $-C$ is proportional to the mass,
and hence $C$ must be negative. From now on we will choose $C=-1$, thus
puting the singularity at $x=0$.

For later conveninece we write the curvature as
\be
R=-{32\over A}\ ,
\ee
where we have introduced
\be
\label{A}
A={8\over\lambda^2}\left(e^{{\lambda^2\over 4}x}-1\right)\ .
\ee

The metric for the general dilaton model, given in terms of $F$ and $\rho$,
is simply
\be
ds^2=e^{2(F+\rho)}(-dt^2+dx^2)\ .
\ee
In the case of CGHS we get
\be
e^{2(F+\rho)}={\lambda^2\over 64}\,
{e^{{\lambda^2\over 4}x}\over e^{{\lambda^2\over 4}x}-1}\ ,
\ee
which vanishes for $x=-\infty$. For stationary metrics the equation
$g_{00}=0$ determines the horizon. Therefore, in these coordinates the CGHS
black hole has a horizon at $x=-\infty$. The curvature, on the other hand,
is well behaved at this point. As with the Schwartzschild black hole one can
now find coordinates which are well behaved at the horizon. In this way one
finally obtains information about the global character of the manifold.

\section{Deformed CGHS Model}

In this section we will construct a new dilaton gravity model that satisfies
the following requrements:
\begin{enumerate}
\item It is completely integrable, i.e. both quadratures can be
solved in closed form.
\item For $x\to\infty$ it goes over into the CGHS model.
\item It is singularity free.
\end{enumerate}

As we have seen, dilaton gravity models are specified by giving the two
potentials $D(\phi)$ and $V(\phi)$. It is very difficult to see how one
should deform these potentials from their CGHS form in order to satisfy the
above criteria. Note, however, that the models are also uniquely determined
by giving $F(\phi)$ and $\tilde V(\tilde\phi)$. This is much better for us
since we have now untangled the two integrability requirements: $F(\phi)$
determines the first quadrature and $\tilde V(\tilde\phi)$ the second.
Deformations of a given model correspond to changes of both of these
functions. In this paper we will look at a simpler problem. We shall keep
$\tilde V(\tilde\phi)$ fixed, i.e. it will have the same value as in
the CGHS model
\be
\tilde V(\tilde\phi)=-{1\over 2}\,\lambda^2\ .
\ee
We will only deform $F(\phi)$. By doing this we are guaranteed that the
second (and more difficult) quadrature is automatically solved. Because of
this $\tilde\phi(x)$ and $\rho(x)$ are the same as in the CGHS model. Using
the value for $\rho(x)$ we may write the scalar curvature for all the
remaining models solely in terms of $F(x)$. We have
\be
\label{R of F}
R=-8\,e^{-{\lambda^2\over 4}x}
\left(e^{-2F}\,{d^2 F\over dx^2}\right)\ .
\ee

Let us now choose $F$. From our second requirement we see that for large $x$
the dilaton field $\phi(x)$ must be near to its CGHS form. Specifically,
$x\to\infty$ corresponds to $\phi\to\infty$. Thus, our second requirement
imposes that for $\phi\to\infty$ we have
\be
F(\phi)\to-\ln\phi\ .
\ee
$F(\phi)$ must also be such that the first quadrature 
(\ref{first quadrature}) is exactly solvable. To do this we choose
\be
\label{F deformed}
F(\phi)=-{1\over\alpha}
\ln\left({1+\beta \phi^\alpha\over\beta}\right)\ ,
\ee
with $\alpha>0$. The $\alpha$ and $\beta$ values parametrize our class of
deformations. The first quadrature now gives
\be
\label{D deform}
D(\phi)=\left\{ 
\begin{array}{ll}
{1\over 8}\phi^2+{1\over 4\beta}\ln\phi&\hbox{  for }\alpha=2\\
{1\over 8}\phi^2+{1\over 4\beta(2-\alpha)}\phi^{2-\alpha}&
\hbox{  for }\alpha\neq 2\ .
\end{array}
\right.
\ee
On the other hand, the potential $V(\phi)$ is now simply
\be
V(\phi)=-{1\over 2}\lambda^2
\left({1+\beta\phi^\alpha\over\beta}\right)^{2\over\alpha}\ . 
\ee

The choice of $\alpha$ corresponds to a choice of explicit model, while
$\beta$ just sets a scale for the dilaton field. Rather than work here with
the general deformed model we will now concentrate on the simplest model in
this class; the one corresponding to the choice $\alpha=4$. The action for
this model is
\be
\label{deformed action}
S=\int d^2x\sqrt{-g}
\left({1\over 2}\,g^{\alpha\beta}\partial_\alpha\phi\partial_\beta\phi
+{1\over 2}\,\lambda^2
\left({1+\beta\phi^4\over\beta}\right)^{1\over 2}
+{1\over 8}\left(\phi^2-{1\over\beta\phi^2}\right)R\right)\ .
\ee

Note that for $\beta\to\infty$ this goes over into the action of the CGHS
model. As we have seen, $\beta$ is just a scale for $\phi$, hence, this is
just a re-statement of our second requirement. From our construction we see
that (\ref{deformed action}) corresponds, for each finite value of $\beta$,
to a model that satisfies our first two requirements. All that is left is
to check that the theory is indeed free of singularities. Being in two
dimensions all that we need to check is the scalar curvature.

From (\ref{phi tilde}) and (\ref{D deform}) for $\alpha = 4$ we find the
connection between
$\phi$ and $\tilde\phi$
\be
\label{relation}
\tilde\phi={1\over 8}\,\left(\phi^2-{1\over\beta\phi^2}\right)\ .
\ee
On the other hand, as we have seen, $\tilde\phi(x)$ is the same as in the
CGHS model, so that (\ref{phi tilde x}) holds. Combining with
(\ref{relation}) we find $\phi^2-{1\over\beta\phi^2}=2A$, where we have
again taken $C=-1$. Equivalently, $\phi^4-2A\phi^2-{1\over\beta}=0$.
This is easily solved --- that is what makes the choice $\alpha=4$ so nice.
We find
\be
\phi^2=A+\sqrt{{1\over\beta}+A^2}\ ,
\ee
where we chose the solution of the quadratic equation that allowed $\phi$ to
go over to $\phi_\mathrm{cghs}$ in the $\beta\to\infty$ limit.

Calculating the scalar curvature is now just a matter of plugging this into
(\ref{R of F}). A simple but tedious calculation now gives
\bea
\label{deformed curvature}
R&=&\sqrt{2}\,\lambda^2\left({1\over\beta}+A^2\right)^{-{7\over 4}}
\left(A+\sqrt{{1\over\beta}+A^2}\,\right)^{1\over 2}\cdot\nonumber \\
& &\qquad\cdot\left\{{16\over\beta\lambda^2}+{3\over\beta}A-
{8\over\lambda^2}A^2+
\left({1\over\beta}-{8\over\lambda^2}A\right)
\sqrt{{1\over\beta}+A^2}\,\right\}\ ,
\eea
where $A(x)$ was given in (\ref{A}). For $\beta\to\infty$ we indeed find that
\be
R\to -{32\over A}\ ,
\ee
which is the CGHS result. From (\ref{deformed curvature}) we see that the
curvature of the deformed CGHS model is indeed not singular.

\begin{figure}[!ht]
  \centering
  \includegraphics[width=10cm]{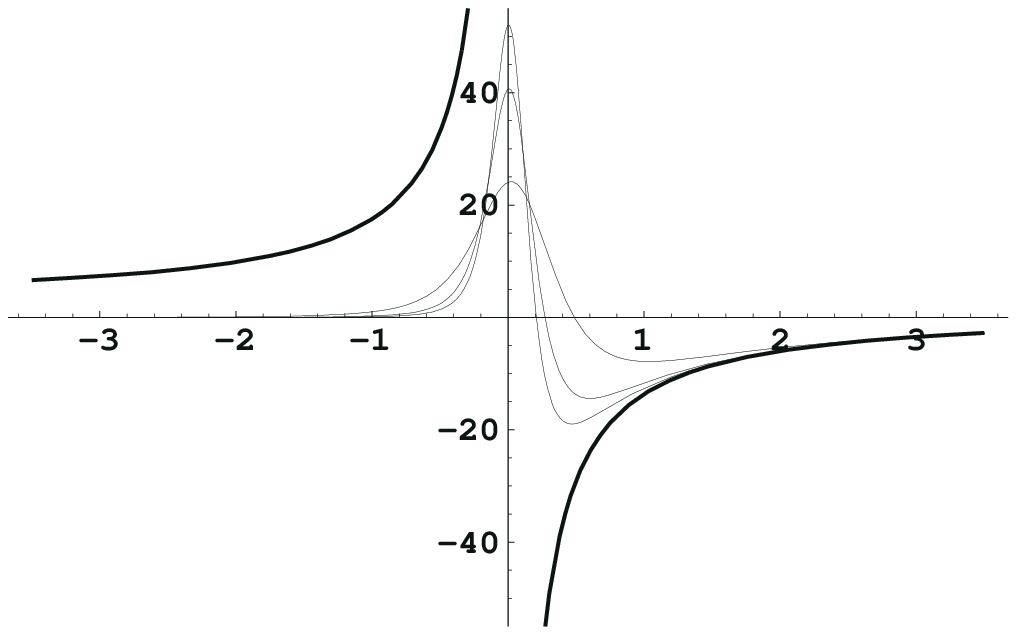}
  \caption{
$R(x)$ for the CGHS model (thick line) and its deformation for $\beta=1$,
3 and 5. As $\beta$ increases the deformations look more and more like the
CGHS result (for $x>0$). The graphs have been plotted for $\lambda^2=1$}
\end{figure}

As may be seen in Figure~1, the deformed model has maximal curvature at
$x=0$. Its value is
\be
\label{R max}
R_\mathrm{max}=\sqrt{2}\,\left(16\beta^{1\over 2}+\lambda^2\right)\ .
\ee
At right infinity the deformed model tends to the CGHS result. On the other
hand, at left infinity both the CGHS model and its deformation tend to a
de Sitter space $R=\Lambda$. However, for CGHS we have $\Lambda=4\lambda^2$,
while for the deformed model the constant is a complicated function of
$\beta$ and $\lambda$. Rather than writing it out let us only give the
result for large $\beta$ when we have
$\Lambda=2^{-10}\lambda^8\beta^{-{3\over 2}}$. 
We have just determined that
\be
\lim_{x\to -\infty}\lim_{\beta\to\infty} R\ne
\lim_{\beta\to\infty}\lim_{x\to -\infty} R\ .
\ee
Put another way: imposing that our model joins to CGHS at right
infinity doesn't automaticaly guarantee a similar joininig at left infinity.

We are now in the position of trying to interpret the meaning of our
deformed CGHS model. Obviously, one possibility is to think of
(\ref{deformed action}) as the classical action
of a model with scale $1\over\beta$. However, it seems more natural to
interpret our model as an effective action. $1\over\beta$ then naturaly
comes about from quantization, while $\beta\to\infty$ corresponds to the
semi classical limit. Our model should thus be the effective action
corresponding to the quantization of the CGHS model. Quantization gives
$S\sim\hbar$, and essentialy dimensional analysis (in units $G=c=1$)
gives $\phi^2\sim\hbar$, as well as ${1\over\beta}\sim\hbar^2$.
Therefore, if we are to interpret our model as an effective action then
$\beta=\kappa\hbar^{-2}$, where $\kappa$ is a constant of the order of
unity. We see then that the maximal curvature (\ref{R max}) is proportional
to ${1\over\hbar}$, i.e. represents a non-perturbative effect.
Expanding our model in $\hbar$ we find
\be
S_\mathrm{eff}=S_\mathrm{cghs}-{1\over 8\kappa}\,\hbar^2\,
\int d^2x\,\sqrt{-g}\,\left(R-2\lambda^2\right)\phi^{-2}+o(\hbar^4)\ .
\label{eff}
\ee
The leading correction to CGHS is of the form of the Jackiw-Teitelboim
action for 2d gravity. It would be very interesting to get this result
by quantizing some fundamental 2d theory. To do this we would need to
start from the CGHS model coupled to some matter fields. We would then
have to integrate out the matter. The last step would be to calculate
the effective action. It is probably impossible to do this exactly,
however, we could hope to do this perturbatively and compare with
(\ref{eff}).

\section{Conclusion}

We have constructed a class of exactly solvable 2d gravity models that
represent deformations of the CGHS dilaton gravity model. In the
semi-classical limit these effective theories go over into the CGHS
model. The deformed CGHS models lead to non-singular black hole
solutions --- i.e. horizons without singularities. We leave a detailed
discussion of the global character of the deformed CGHS solutions for
a later publication.

It will be intersting to apply this method to non-singular 2d cosmology
models. A further avenue of research is to consider deformed models for
dilaton gravity in the presence of matter.

\end{document}